\documentclass[12pt, a4paper,fleqn]{article}
\usepackage{listings}
\usepackage{pdfpages}
\usepackage{subfigure}
\usepackage{stmaryrd}
\usepackage{txfonts}
\usepackage{appendix}

\usepackage{amsmath}
\usepackage{bbm}
\usepackage{titlesec}
\usepackage{amsfonts}
\usepackage{float}
\usepackage{amssymb}
\usepackage{amsmath,amsfonts,latexsym}
\usepackage{graphicx}

\newcommand{\eps}{\varepsilon}  
\title{\textbf{Separatrix crossing in rotation of a body with changing geometry of masses}}
\author{Jinrong Bao$^1$, Anatoly  Neishtadt$^{1,2}$
\footnote{Corresponding author.  E-mail addresses: {\it J.Bao@lboro.ac.uk (J.Bao)} {\it A.Neishtadt@lboro.ac.uk} (A.Neishtadt)} \\ 
$^1$ Loughborough University, Loughborough, LE11 3TU, UK\\
$^2$ Space Research Institute, Moscow, 117997, Russia}
\date{ }
\begin{document}
\maketitle
\begin{abstract}
We consider free rotation of a body whose parts move slowly with respect to each other under the action of internal forces.  This problem can be considered as a perturbation of the Euler-Poinsot problem. The dynamics has an approximate conservation law - an adiabatic invariant. This allows to describe the evolution of rotation in the adiabatic approximation. The evolution leads to an overturn in the rotation of the  body: the vector of angular velocity crosses the separatrix of the Euler-Poinsot problem.   This  crossing leads to a quasi-random scattering in body's dynamics. We obtain formulas for probabilities of capture into different domains in the phase space at separatrix crossings.  
\end{abstract}
\section*{Introduction}
Rotation of a rigid body around its centre of mass is a classical problem in mechanics. The case when parts of the body slowly move with respect to each other can be considered as a perturbation to the case of a rigid body.  This problem  serves  as a model of rotation of deformable celestial bodies (see, e.g., \cite{munk, goldreich}).   We consider  a free rotation of a body with  a prescribed slow motion of parts of the body with respect to each other. This problem  is a perturbation of the Euler-Poinsot problem. We use averaging method and adiabatic approximation  to give a description of the rotational motion of a  body. The probabilistic part of description shows the probabilities of overturns of the body at a crossing of the separatrix of Euler-Poinsot problem.

\section{Description of the system and equations of motion}
\begin{figure}[H]
\centering
\includegraphics[width=0.6\textwidth]{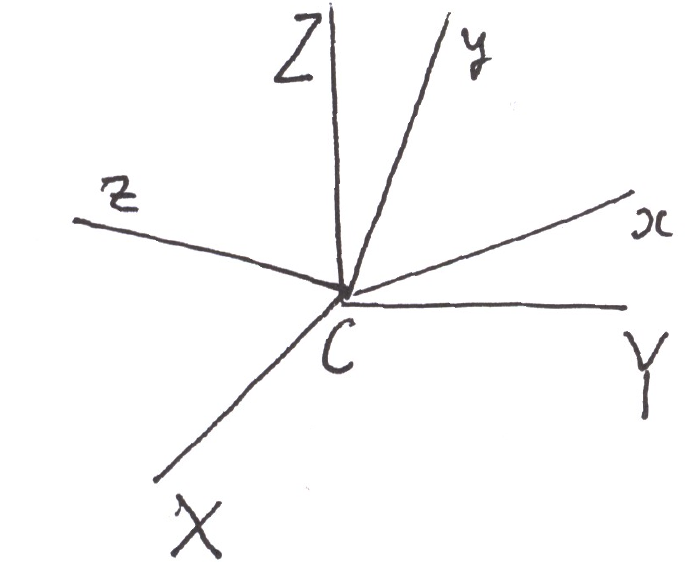}
\caption{K\"onig's frame and body frame.}
\label{frames}
\end{figure}
Consider motion of a system of particles. We assume that there are no external forces acting on these particles. Then the centre of mass ${\cal {C}}$ of this system moves with a constant velocity and the angular momentum of this system about  ${\cal {C}}$ is constant throughout the motion.
% \cite{arnol2013mathematical}. 
Consider two frames of reference: a non-rotating frame ${\cal {C}}XYZ$ (K\"onig's  frame)  and the frame ${\cal {C}}xyz$ whose  axes  are  principal axes of inertia  of the system of  particles, Fig. \ref{frames}.
% In what follows without loss of generality we assume that ${\cal {C}}$ does not move. Thus ${\cal {C}}XYZ$ can be considered as an absolute frame of reference.
We assume that the motion of the particles with respect to each other is prescribed in advance. An example is the motion of a rigid body and particles which move in a prescribed way with respect to this body. (In this example the rigid body is considered as a system of particles with fixed distances with respect to each other.)  One can consider also an object which consists of several rigid bodies moving with respect to each other.  The frame ${\cal {C}}xyz$ rotates with some angular velocity. Denote $\overrightarrow{\omega}$  this  angular velocity considered as a vector in the frame  ${\cal {C}}xyz$. Denote $\overrightarrow{G}$ the angular momentum of the moving particles with respect to the point $C$;  $\overrightarrow{G}$ also is considered as a vector in the frame  ${\cal {C}}xyz$. Derive, following \cite{routh}, Ch. 1,  equations of motion of the vector  $\overrightarrow{G}$.  Denote $m_i$  masses of particles  and $\overrightarrow{r_i}$ their position vectors in the frame ${\cal {C}}xyz$.  The particle number $i$ moves with respect to the frame ${\cal {C}}xyz$, and its relative velocity is $\overrightarrow{v_i}=\dot{\overrightarrow{r_i}}$. Notice that in the absolute space this particle also moves due to rotation of the frame ${\cal {C}}xyz$. This implies that its  velocity with respect to the frame ${\cal {C}}XYZ$ is $\overrightarrow{V_i}=\overrightarrow{\omega}\times\overrightarrow{r_i}+\overrightarrow{v_i}$. Then by the definition of the angular momentum  about $C$ we have
\begin{equation}
\overrightarrow{G}=\sum_i m_i\overrightarrow{r_i}\times\overrightarrow{V_i}=\sum_i m_i\overrightarrow{r_i}\times(\overrightarrow{\omega}\times\overrightarrow{r_i})+\sum_i m_i\overrightarrow{r_i}\times\overrightarrow{v_i}=\hat{I}\overrightarrow{\omega}+\overrightarrow{g}.
\end{equation}
Here $\hat{I}=\hat{I}(t)$ is the matrix of the inertia tensor of the particles in the axes ${\cal {C}}xyz$,   $\overrightarrow{g}=\overrightarrow{g}(t)$ is the angular momentum about $C$ of the  particles in the motion with respect to the frame ${\cal {C}}xyz$. Then
$$\overrightarrow{\omega}=\hat{I}^{-1}\overrightarrow{G}-\hat{I}^{-1}\overrightarrow{g}.$$
Denote $\overrightarrow f(t)=\hat{I}^{-1}\overrightarrow{g}$. Then
$$\overrightarrow{\omega}=\hat{I}^{-1}\overrightarrow{G}-\overrightarrow f(t).$$
The conservation of the total angular momentum of the system  in K\"onig's frame implies that
\begin{equation}\label{euler}
\frac{d\overrightarrow{G}}{dt}+\overrightarrow{\omega}\times\overrightarrow{G}=0.
\end{equation}
Plugging  $\overrightarrow{\omega}$ into the equation above, we get the following equation known as Liouville's equation (see, e.g., \cite {munk}) 
\begin{equation}\label{Euler's}
\frac{d\overrightarrow{G}}{dt}+(\hat{I}^{-1}\overrightarrow{G}-\overrightarrow f(t))\times\overrightarrow{G}=0.
\end{equation}
In this equation $\hat{I}$ and $\overrightarrow f$ are prescribed functions of time. We will  assume that the motion of particles is slow, of order $\eps$, where $\eps>0$ is the small parameter of the problem.  Then  changes of   $\hat{I}$ and $\overrightarrow f$ with time are slow, of order $\eps$, and  $\overrightarrow f$ is small, of order $\eps$. So, with a slight change of notation ($\hat{I}(t) \to \hat{I}(\tau),\overrightarrow  f(t) \to \eps \overrightarrow f(\tau)$), we rewrite equation (\ref{Euler's}) as 
\begin{equation}\label{Euler's_1}
\frac{d\overrightarrow{G}}{dt}+(\hat{I}^{-1}(\tau) \overrightarrow{G}- \eps \overrightarrow f(\tau))\times\overrightarrow{G}=0, \quad \dot \tau=\eps.
\end{equation}
This system with $f(\tau)\equiv 0$ was studied in \cite{Bor_Mam}.

\section{Unperturbed system}

\begin{figure}[H]
\centering
\includegraphics[width=0.6\textwidth]{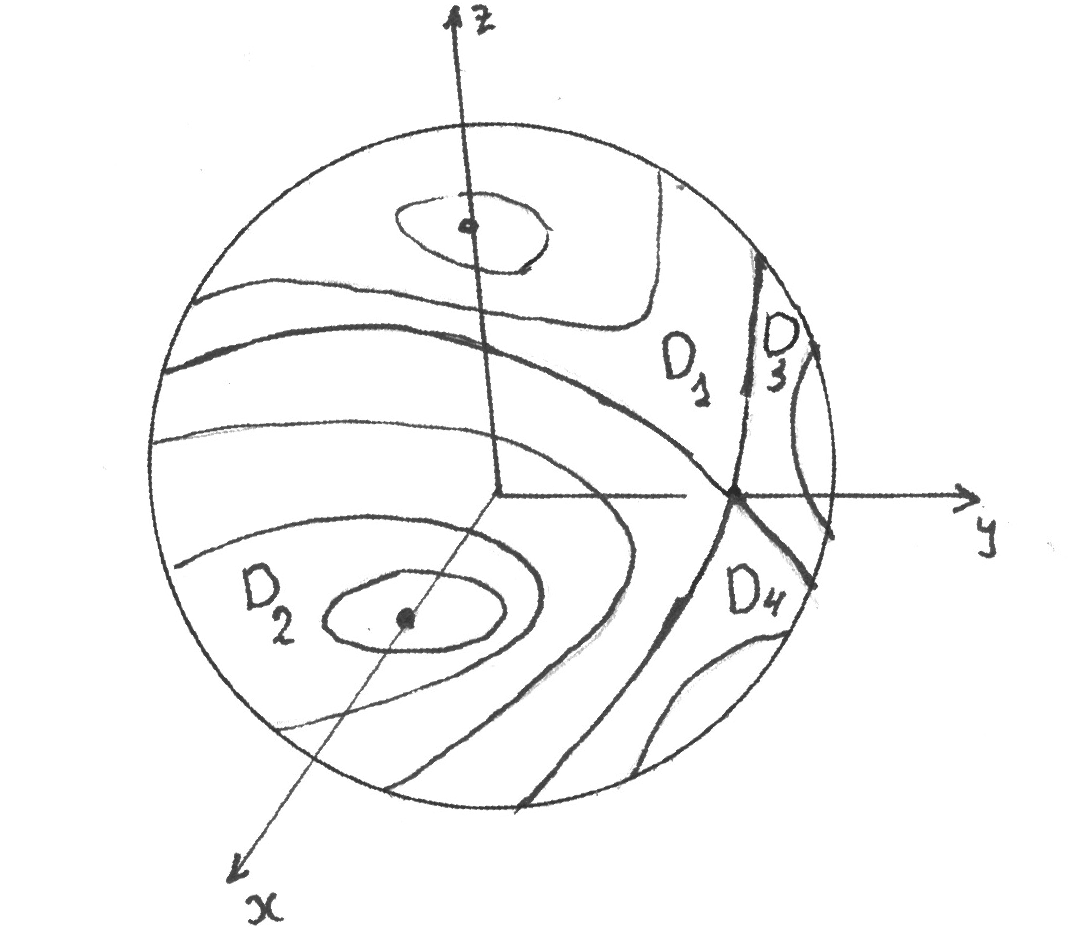}
\caption{Rigid body dynamics on the angular momentum sphere.}
\label{sphere}
\end{figure}

\begin{figure}[H]
\centering
\includegraphics[width=0.55\textwidth]{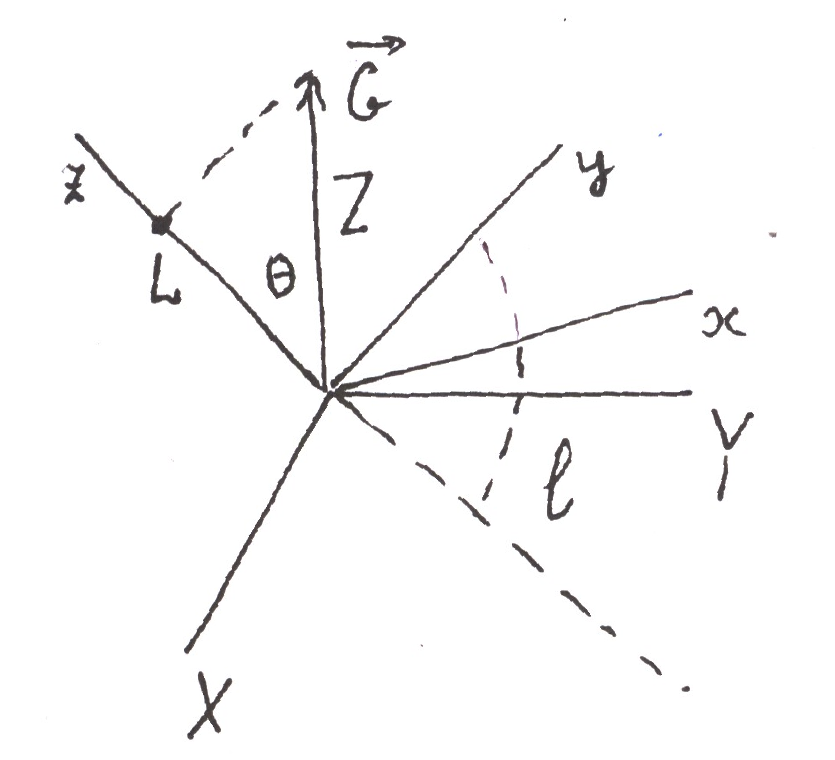}
\caption{Andoyer-Deprit variables.}
\label{AD_variables}
\end{figure}

Consider the unperturbed system, i.e. system (\ref{Euler's_1}) for $\eps=0, \tau={\rm const}$. We get Euler-Poinsot problem. It can be considered in several representations: on Poinsot ellipsoid (ellipsoid of a constant kinetic  energy), on sphere of a constant magnitude  of the angular momentum vector, 
% $|\vec G|={\rm const}$,
 and in the phase plane in  Andoyer-Deprit variables. We will use the last two  representations.  Denote $A$, $B$, and $C$ the principal moments of inertia of the body, $A$ corresponds to ${\cal {C}}x$ axis, etc.  We assume that principal moments of inertia are prescribed functions of the parameter  $\tau$ ($A=A(\tau)$, etc.), and their values never coincide.  Without loss of generality, assume that $A>B>C$. Vectors  $\vec \omega =(\omega_1, \omega_2, \omega_3)$ and $\vec G=(G_1, G_2, G_3)$ are related as $\vec\omega=  (\frac{G_1}{A}, \frac{G_2}{B},\frac{G_3}{C})$.  
 %Poinsot ellipsoid is determined by the equation
%$$
%\frac12\left (A\omega_1^2+B\omega_2^2+C\omega_3^2\right)=h
%$$
%where $h={\rm const}$  is the kinetic energy of the body. Trajectories of the vector $\vec \omega$ on this ellipsoid are isolines of the length of the  of the angular momentum vector (Fig. :)
%$$
%A^2\omega_1^2+B^2\omega_2^2+C^2\omega_3^2=G^2. 
%$$
The  sphere  $|\vec G|={\rm const}$ in coordinates  $(G_1, G_2, G_3)$ has the equation
$$
G_1^2+G_2^2+G_3^2=G^2
$$
where $G=|\vec G|$. 
Trajectories of the vector $\overrightarrow G$ on this sphere (see Fig. \ref{sphere}) are isolines of kinetic energy  $H_0$ of rotation of the body,
$$
H_0= \frac12\left (\frac {G_1^2}{A}+\frac {G_2^2}{B}+\frac{G_3^2}{C}\right).
$$
Let us take direction of the vector $\vec G$ as the positive direction of $CZ$ axis (note that  $\vec G$ is a constant vector in the absolute space).  We will use   Andoyer-Deprit variables $L,l$, where $L$ is  the projection of $\overrightarrow{G}$ onto the axis corresponding to the moment of inertia $C$,  and $l$ is  the intrinsic rotation angle of the body (Fig. \ref{AD_variables} ).  
These variables are related to the components of the angular momentum vector as follows:
\begin{equation}
\label{G_L_relation}
G_1=G\sin{\theta}\sin{l},\
G_2=G\sin{\theta}\cos{l},\
G_3=L,
\end{equation}
where $\sin{\theta}=\frac{\sqrt{G^2-L^2}}{G}$.\\
Dynamics in Euler-Poinsot problem in Andoyer-Deprit  variables  is described by the Hamiltonian system with one degree of freedom, where $l,L$ are conjugate canonical variables \cite{deprit1967free}. Kinetic energy $H_0$ is the Hamiltonian: 
\begin{equation}
\label{f_Hamiltonian}
H_0=\frac{1}{2}(\frac{\sin^2{l}}{A}+\frac{\cos^2{l}}{B})(G^2-L^2)+\frac{L^2}{2C}.
\end{equation}

\begin{figure}[H]
\centering
\includegraphics[width=0.7\textwidth, angle=-1.5]{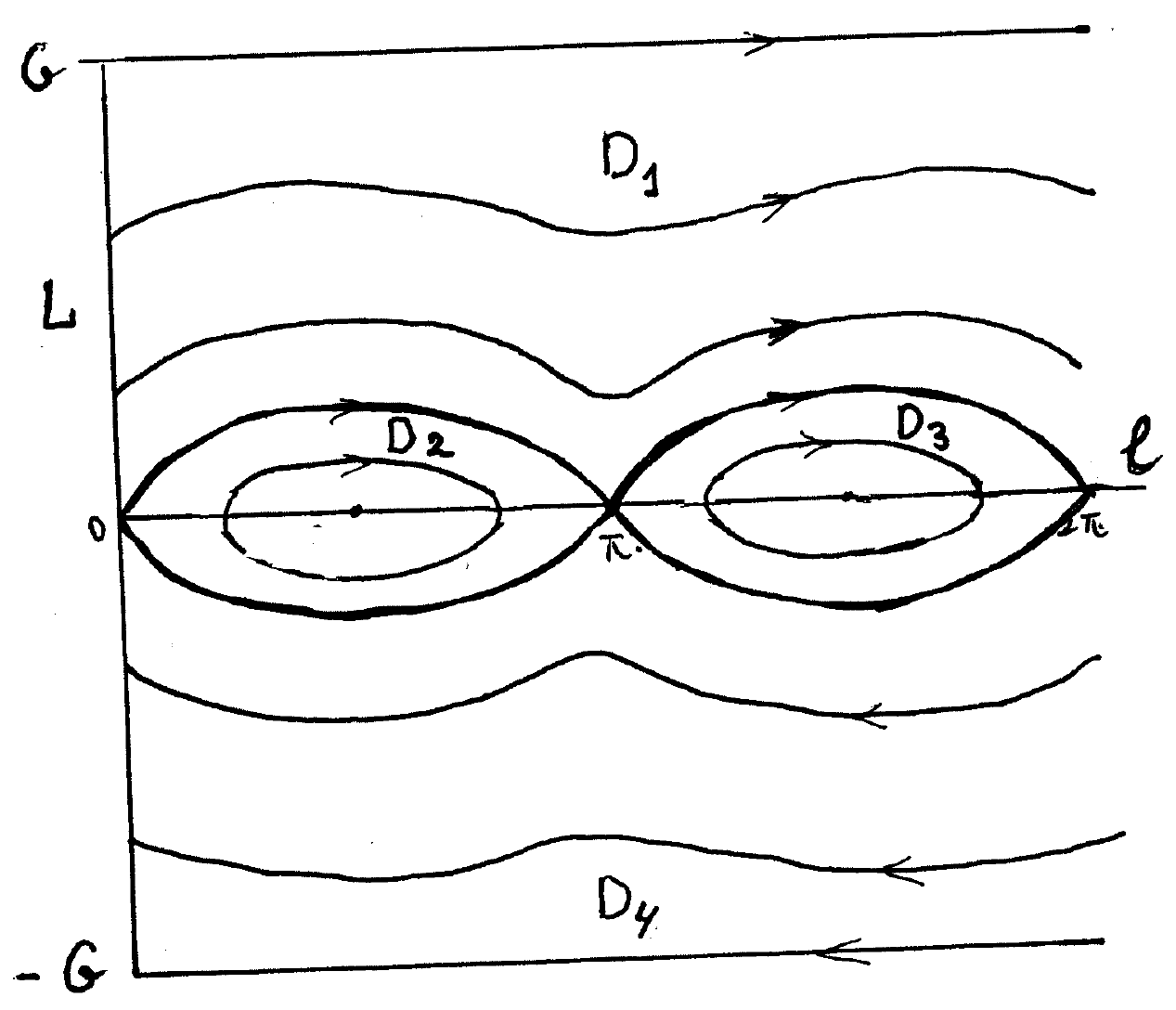}
\caption{Phase portrait of the Euler-Poinsot problem.}
\label{rectangle}
\end{figure}

The equations of motion of the system are
$$\dot{l}=\frac{\partial{H_0}}{\partial{L}}=(\frac{1}{C}-\frac{\sin^2{l}}{A}-\frac{\cos^2{l}}{B})L\,, \quad
\dot{L}=-\frac{\partial{H_0}}{\partial{l}}=\frac{1}{2}(\frac{1}{B}-\frac{1}{A})(G^2-L^2)\sin{2l}.$$
The phase portrait of this system should be considered in the cylinder $\{  l \ {\rm mod}\ 2\pi,\ L\}$ (Deprit cylinder). It is shown in the rectangle $\{ 0\le  l\le  2\pi,\  -G\le L \le G  \}$ in Fig. \ref{rectangle}.
% \ref{rectangle}. 

%The phase portrait of this system in the plane $L,l$ (the Deprit plane) is shown in Fig. \ref{AD_plane}. 

%Note that  value $G$ is conserved both in unperturbed and perturbed problems. 

Stable equilibria in this phase portrait  correspond to stationary rotations about ${\cal {C}}x$-axis in positive (for  $l=\pi/2$) and negative   (for  $l=3\pi/2$) directions. Unstable equilibria correspond to stationary rotations about ${\cal {C}}y$-axis in positive (for  $l=0$) and negative   (for  $l=\pi$) directions. Horizontal lines  $L=G$ and $L=-G$ correspond to stationary rotations about ${\cal {C}}z$-axis in positive and negative directions, respectively.  Values of kinetic energy for stationary rotations about ${\cal {C}}x, {\cal {C}}y$ and ${\cal {C}}z$ axes are $\frac {G^2}{2A}, \frac {G^2}{2B}$ and $\frac {G^2}{2C}$, respectively.

Separatrices divide the phase portrait into domains $D_1, D_2, D_3, D_4$ (see Fig. 4). These domains correspond to domains  with the same names on the sphere of the constant angular momentum, Fig. \ref {sphere}. 
%We denote $\Gamma_{i,j}$ the separatrix which separates domains $D_i$ and $D_j$. 
Separatrices are determined by the equation
$$\frac{1}{2}\left(\frac{\sin^2{l}}{A}+\frac{\cos^2{l}}{B}\right)(G^2-L^2)+\frac{L^2}{2C}=\frac{G^2}{2B}$$
which simplifies to
\begin{equation}
\label{separatrix_eq}
L=\pm G\sin{l}\left({\frac{\frac{1}{C}-\frac{1}{A}}{\frac{1}{B}-\frac{1}{A}}-\cos^2{l}}\right)^{-1/2}.$$
\end {equation}

Action variable $I=I(h, \tau)$ in this phase portrait is defined separately in each domain. In domains  $D_1$ and $D_4$ this is the area between the line $H_0=h$ and the line $L=0$ for $0\le l\le 2\pi$, divided by $2\pi$. In domains  $D_2$ and $D_3$ this is the area surrounded by the  line $H_0=h$, divided by $2\pi$. The formula for  $I=I(h, \tau)$  is (see \cite{sadov1970action}, a misprint in this paper is corrected in  \cite{lara2010integration})
\begin{equation}
\label{I_Sadov}
I(h, \tau)=\frac{2G}{\pi\kappa}\left(\frac{1+\kappa^2}{\lambda+\kappa^2}\right)^{\frac{1}{2}}[(\lambda+\kappa^2)\Pi(\frac{\pi }{2},\kappa^2,\lambda)-\lambda K(\lambda)],
\end{equation}
where $\kappa$ and $\lambda$ are positive parameters given by
$$\kappa^2=\frac{C(A-B)}{A(B-C)},\quad \lambda=\frac{(A-B)(G^2-2Ch)}{(B-C)(2Ah-G^2)},$$
$K$ is the complete elliptic integral of the first kind, $\Pi$ is an elliptic integral of the third kind.

In what follows we need formulas for areas of domains $D_i$. They can be obtained  as limiting values of $2\pi I(h, \tau)$ as $h \to  \frac {G^2}{2B}$, or by integrating $L$ given by equation (\ref{separatrix_eq}), or geometrically, from areas on angular momentum sphere. Either way leads to the following results. Area of each of domains $D_2$, $D_3$ is
\begin{equation}\label{area}
S=4G\arcsin{k}, \quad  k = \sqrt{ \frac{\frac{1}{B}-\frac{1}{A}}{\frac{1}{C}-\frac{1}{A}} }.
\end{equation} 
Area of each of domains $D_1$, $D_4$ is 
\begin{equation}\label{area1}
\tilde S=2\pi G - S=2G(\pi-2\arcsin{k}).  
  \end{equation} 
  Areas of the corresponding domains on angular momentum sphere  are   $GS$ and $G\tilde S$.

\section{Adiabatic approximation for perturbed system}
\label{ad_approximation}
Now consider system (\ref{Euler's_1}) for $\eps>0,\  \dot \tau=\eps$. We will call $\tau$  ``slow time" or just ``time" when this does not lead to a mixup.  Dynamics of Andoyer-Deprit variables is described by the Hamiltonian system  with the Hamilton's function
\begin {equation}
\label{H_general}
H=H(l,L,\tau)=H_0(l,L,\tau)+\eps H_1(l,L,\tau).
\end {equation} 
Here $H_0$ is the kinetic energy of the body (see (\ref {f_Hamiltonian})). The function $H_1$ is given by the formula 
%(see Appendix 1)
\begin {equation}
\label{f_H1}
H_1= (\vec f\cdot\vec G)=f_1\sqrt{G^2-L^2}\sin l+f_2\sqrt{G^2-L^2}\cos l+f_3L
\end {equation}
where $f_i, \, i=1,2,3$ are components of the vector $\vec f$.

Dynamics of Hamiltonian systems with slowly varying parameters of the form (\ref{H_general}) can be described in an adiabatic approximation (see, e.g.,\cite{arnold2007mathematical}, Sect. 6.4). In particular, the value of the action  (\ref{I_Sadov}) is approximately conserved in the process of motion. The adiabatic approximation in this problem was used in \cite{goldreich} for description of dynamics far from separatrices on the sphere of the angular momentum.  

It may happen that in the process of the perturbed motion the phase point in the Deprit cylinder crosses a separatrix of the unperturbed system.  Initially, at $t=0$, this phase point is in a domain $D_i$, but after some time of order $1/\eps$ it moves in a domain $D_j$, $j\ne i$.

Adiabatic approximation can be used up to the separatrix \cite{N2}. Thus the moment of separatrix crossing  can be approximately found from conservation of the adiabatic invariant as follows. Let $I_0$ be initial, at $\tau=0$, value of action. Areas of domains $D_{2,3}$ and $D_{1,4}$ are functions of $\tau$: $S(\tau)= 4G\arcsin k(\tau)$ and $\tilde S(\tau)= 2G(\pi - 2\arcsin k(\tau))$, respectively. Then the moment $\tau_*$ of separatrix crossing is calculated  in the adiabatic approximation as the closest to $\tau=0$ root of the equation $S(\tau)=2\pi I_0$. Separatrix crossing leads to an overturn of the body. 

\section{Separatrix crossing}
\subsection{Separatrix crossing in adiabatic approximation}
Consider motion of phase points which initially, at $\tau=0$, are in $D_1$ at a distance of order 1  from separatrices.    We assume that $dS/d\tau > {\rm const}>0$. Thus areas of domains $D_2$ and $D_3$ grow, areas of domains $D_1$ and $D_4$ decay.   The phase points approach  separatrices at $\tau\approx \tau_*$,  where  $\tau_*$ is the moment of separatrix crossing calculated in the adiabatic approximation Sect. \ref{ad_approximation}.   We are interested in further motions of these phase points. The averaging method for dynamics with separatrix crossings (see, e.g., \cite{N2} and references therein) allows to describe this motion in the adiabatic approximation as follows.   At the separatrix crossing the phase point can not continue its  motion in the domain $D_4$ because area of the domain $D_2\cup D_3$  grows.  The phase point continues its motion either in the  domain $D_2$ or in the domain $D_3$. In each case the value of the adiabatic invariant $I$ is equal to $S(\tau_*)/(2\pi)$.  However $S(\tau_*)=2\pi I_0$,  where $I_0$ is initial, at $\tau=0$, value of action. Thus $I=I_0$ in the adiabatic approximation for motions with separatrix crossings.

One can assign certain probabilities to continuations of motion in $D_2$ and $D_3$. We will calculate these probabilities in Sect.  \ref{ss_prob}.

\subsection{Change of energy near separatrices}
In order to study separatrix crossing we need to know asymptotic formulas for change of energy $H_0$ in the perturbed system when the phase point moves near unperturbed separatrices (cf. \cite{N2}).  The energy along the separatrices is $H_B=\frac{G^2}{2B}$.  Introduce the new function $E =H-H_B$.  In the perturbed system (\ref{Euler's_1}) we have
$$
  \frac{dE}{dt}=\varepsilon\left(\frac{\partial{E}}{\partial{\tau}} + \nabla{H}\cdot (\vec f\times\vec  G)\right).
  $$
Here  $\nabla{H}$ is the gradient of $H$ considered as the  function of $G_1, G_2, G_3$:
  $\nabla{H}=(G_1/A, G_2/B, G_3/C)$.
  
\begin{figure}[H]
\centering
\includegraphics[width=0.8\textwidth,height=0.4\textheight]{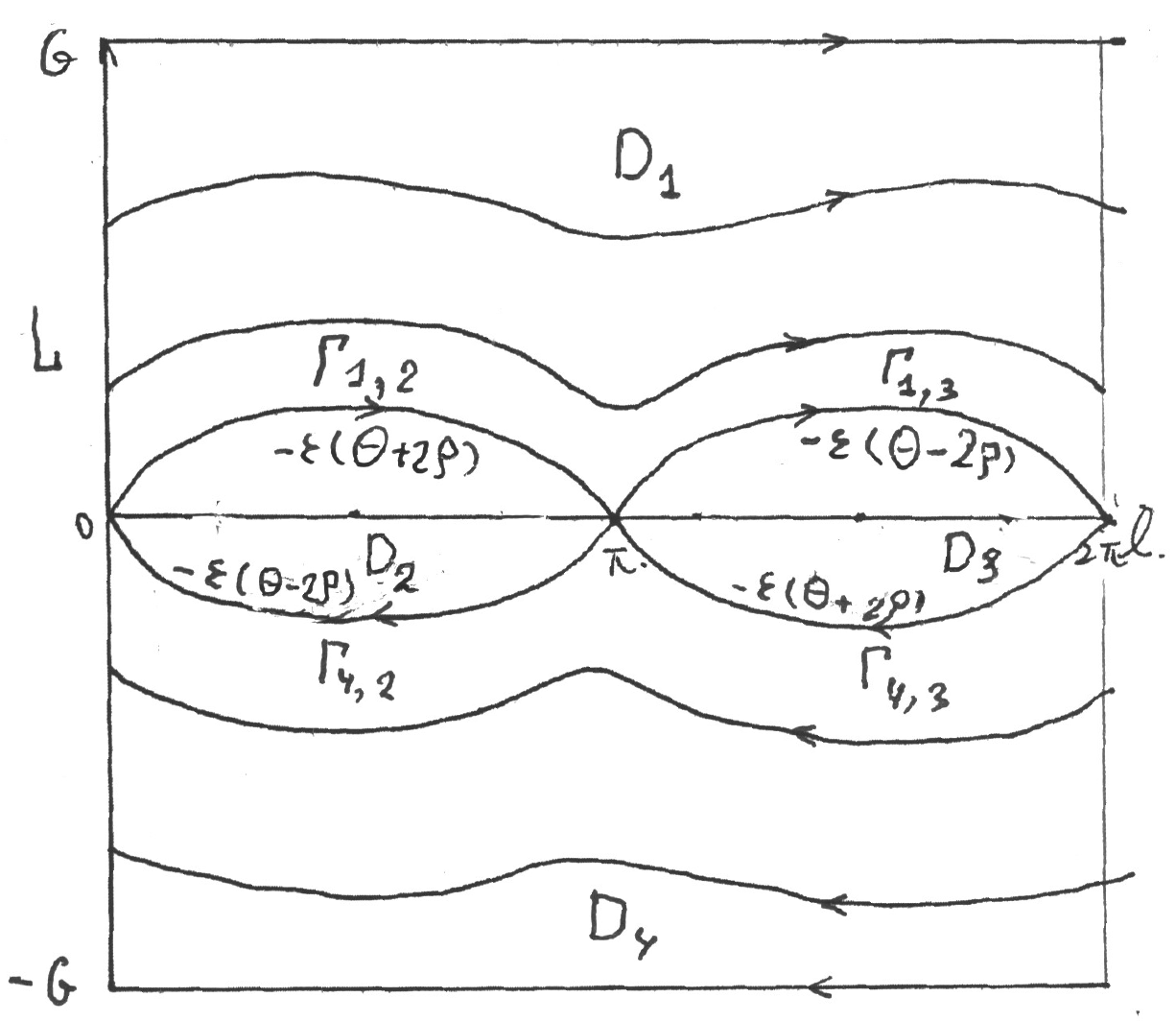}
\caption{Separatrices.}
\label{crossing}
\end{figure}

  The separatrices  divide the phase cylinder into four parts: $D_1$, $D_2$, $D_3$, and $D_4$. There are four separatrices, see Fig. \ref {crossing}. Denote the separatrix between $D_1$ and $D_2$ as $\Gamma_{1,2}$. Similarly we have $\Gamma_{1,3}$, $\Gamma_{4,2}$ and $\Gamma_{4,3}$.
\ \\
 Change $\Delta E$ of value of the function $E$ along a fragment of perturbed trajectory close to a separatrix $\Gamma$ is approximately equal  to the integral of  $dE/dt$ over $\Gamma$:
 \begin{equation}
 \label{two_int}
  \Delta E \approx \varepsilon\int_{\Gamma}\left(  \frac{\partial{E}}{\partial{\tau}}\right)dt  + \varepsilon \int_{\Gamma}\nabla{H}\cdot (\vec f\times \vec G) dt .
  \end{equation}
  Estimates of accuracy of such an approximation are contained in \cite{N2}. In particular, for motion with $E\sim\eps$ the accuracy is $O({\eps}^{3/2})$. 
 The integrals here are  improper ones, as the motion along a separatrix takes infinite time, but they converge.
\ \\
For calculation of the first integral in  (\ref{two_int}) we use the formula that the $\tau$-derivative of the area, surrounded by separatrices,  is equal to the integral of  $-(\partial E/\partial \tau)$ over these separatrices (cf. \cite {N1}). Denote 
$$\Theta=\Theta(\tau) =\frac12 \frac {\partial{S}}{\partial{\tau}},$$
 where  $S$ is the area of each of domains $D_{2,3}$. Thus 
 the first integral in  (\ref{two_int}) is
$$
 \int_{\Gamma}\left(  \frac{\partial{E}}{\partial{\tau}}\right)dt =-\Theta= -\frac12\frac{\partial{S}}{\partial{\tau}}.
$$
for each of separatrices.
From Sect. 2 we know that  $S=4G\arcsin{k}$ with $k^2=\frac{\frac{1}{A}-\frac{1}{B}}{\frac{1}{A}-\frac{1}{C}}$. Then
\begin{equation}
\begin{split}
\frac{\partial{S}}{\partial{\tau}}&=4\frac{\partial{G\arcsin{k}}}{\partial{\tau}}
    =\frac{4G}{\sqrt{1-k^2}}\frac{dk}{d\tau}
    =\frac{4G}{\sqrt{1-k^2}}\frac{d}{d\tau}\sqrt{\frac{\frac{1}{A}-\frac{1}{B}}{\frac{1}{A}-\frac{1}{C}}}\\
    &=2G\sqrt{\frac{\frac{1}{A}-\frac{1}{C}}{\frac{1}{B}-\frac{1}{C}}}\sqrt{\frac{\frac{1}{A}-\frac{1}{C}}{\frac{1}{A}-\frac{1}{B}}}
    \cdot\frac{\left(\frac{A'(\tau)}{A^2}-\frac{B'(\tau)}{B^2}\right)\left(\frac{1}{A}-\frac{1}{C}\right)-\left(\frac{1}{A}-\frac{1}{B}\right)\left(\frac{A'(\tau)}{A^2}-\frac{C'(\tau)}{C^2}\right)}{\left(\frac{1}{A}-\frac{1}{C}\right)^2}\\
    &=2G\frac{\left(\frac{A'(\tau)}{A^2}-\frac{B'(\tau)}{B^2}\right)\left(\frac{1}{A}-\frac{1}{C}\right)-\left(\frac{1}{A}-\frac{1}{B}\right)\left(\frac{A'(\tau)}{A^2}-\frac{C'(\tau)}{C^2}\right)}{\left(\frac{1}{A}-\frac{1}{C}\right)\sqrt{\left(\frac{1}{B}-\frac{1}{C}\right)\left(\frac{1}{A}-\frac{1}{B}\right)}}.\\
\end{split}
\nonumber
\end{equation}
 Therefore,
$$\int_{\Gamma}\frac{\partial{E}}{\partial{\tau}}dt=-\Theta=- \frac{1}{2}\frac{\partial{S}}{\partial{\tau}}=-G\frac{\left(\frac{A'(\tau)}{A^2}-\frac{B'(\tau)}{B^2}\right)\left(\frac{1}{A}-\frac{1}{C}\right)-\left(\frac{1}{A}-\frac{1}{B}\right)\left(\frac{A'(\tau)}{A^2}-\frac{C'(\tau)}{C^2}\right)}{\left(\frac{1}{A}-\frac{1}{C}\right)\sqrt{\left(\frac{1}{B}-\frac{1}{C}\right)\left(\frac{1}{A}-\frac{1}{B}\right)}}.$$
Here ``prime'' denotes the derivative with respect to $\tau$.
 
\ \\
Now we compute the second integral in equation (\ref{two_int}). We have
$$
\nabla{H}\cdot (\vec f\times \vec G)=\frac{\partial{H}}{\partial{\overrightarrow{G}}}\cdot(\vec f\times\overrightarrow{G})=(\overrightarrow{G}\times\frac{\partial{H}}{\partial{\overrightarrow{G}}})\cdot \vec f=(\overrightarrow{G}\times\hat{I}^{-1}\overrightarrow{G})\cdot \vec f=\frac{d\overrightarrow{G}}{dt}\cdot \vec f,$$
where $\frac{d\overrightarrow{G}}{dt}$ is calculated in the unperturbed system.
\ \\
We know that the point $l=0, L=0$ in the  phase portrait Fig. \ref{crossing} corresponds to a rotation around the axis of the moment of inertia $B$ in the positive direction ($G_2=G$), while the point $l=\pi, L=0$ corresponds to a rotation around this axis  in the negative direction ($G_2=-G$). Therefore,
\begin{equation}\label{2nd}
\int_{\Gamma_{1,2}}\nabla{H}\cdot (\vec f\times \vec G)dt=\int_{\Gamma_{1,2}}\frac{d\overrightarrow{G}}{dt}\cdot \vec fdt=\vec f\cdot\int^{\pi}_{0}d\overrightarrow{G}=-2 f_2G,
\end{equation}
where $f_2$ is the second component of $\vec f$. Integral over  $\Gamma_{4,3}$ has the same value $-2 f_2G$, integrals over $\Gamma_{1,3}$ and $\Gamma_{ 4,2}$ have value $2 f_2G$.
\ \\
Denote $\rho=\rho(\tau)=f_2G.$ Then from formula (\ref{two_int}) and  previous calculations we have the following  expressions for change $\Delta_{i,j} E$ of value of the function $E$ along a fragment of perturbed trajectory close to a separatrix $\Gamma_{i,j}$, valid in the principal approximation:
\begin{eqnarray}
\label{p_delta}
\Delta_{1,2} E= \varepsilon(-\Theta - 2\rho),\quad
\Delta_{1,3} E= \varepsilon(-\Theta + 2\rho),\\
\Delta_{4,2} E= \varepsilon(-\Theta + 2\rho), \quad
\Delta_{4,3} E= \varepsilon(-\Theta - 2\rho) \nonumber.
\end{eqnarray}
These values of change of the function $E$ for motion near the separatrices are indicated in Fig. \ref{crossing}.

Denote $ \Delta_1$ and  $ \Delta_4$  changes of $E$ along  fragments of perturbed trajectory $O(\eps)$-close to $\Gamma_1=\Gamma_{1,2}\cup\Gamma_{1,3}$ and to  $\Gamma_4=\Gamma_{4,3}\cup\Gamma_{4,2}$, respectively.  We have the following expression valid in the adiabatic approximation: 
\begin{eqnarray*}
\Delta_1= \Delta_4=-2\varepsilon\Theta\,.
\end{eqnarray*}

In what follows we use these approximate formulas as exact formulas for change of energy, thus neglecting high order corrections.

\subsection{Probabilities of capture into different domains}
\label{ss_prob}
We consider phase points that start their motion at $\tau=0$ in $D_1$.  These phase points can be captured into $D_2$ or $D_3$ after separatrix crossing. Initial conditions for capture into different domains are mixed in the phase space. Small, $\sim \eps$, change of initial conditions can change the destination of a phase point after separatrix crossings.  Destinations depend  also on $\eps$.  For a fixed initial condition and $\eps \to 0$ two destinations replace each in an oscillatory manner. Thus the question about this destination does not have a deterministic answer in the limit as $\eps \to 0$.    It is reasonable to consider capture into a given domain as a random event and calculate the probability of this event. Such an approach was introduced in \cite {LSN} for  scattering of charged particles at separatrix crossing. This approach was rediscovered by many authors. In particular, it was used in \cite{GP}  for  planetary rotations  to  determine the probability of capture of Mercury in its current resonant regime.  A rigorous approach to the definition of the notion of probability in the considered class of problems was suggested in \cite {A63}. It is based on the comparison of measures of initial data for captures into different domains. Below we use this  approach.

Consider a point $  M$ in the domain $D_1$ at a distance of order 1 from separatrices.  Denote $ I_M$ the value of action at $ M$ at $\tau=0$.   Let    $\tau_*$ be the moment of separatrix crossing  in adiabatic approximation for motion with the initial condition at    $  M$, $S( \tau_*)=2\pi   I_M$.  We will consider motion in the time interval $0\le\tau\le K$, where $K={\rm const}>  \tau_*$. Thus, this interval includes the moment of separatrix crossing (in the adiabatic approximation) for the motion with the initial condition at $  M$.

  Denote $U^{\delta}$ the  disc of radius $\delta$ with the centre at  $  M$. We assume that $\delta$ is small enough, and thus $U^{\delta}$ is in $D_1$ at $\tau=0$. Moreover, the moment of separatrix crossing (in the adiabatic approximation) for all motions with initial conditions in $U^{\delta}$ is less than $K$.   The set $U^{\delta}$ can be represented as a union of three sets,  $U^{\delta}=U^{\delta,\eps}_1\cup U^{\delta,\eps}_2\cup v_{\eps}$ as follows. The set $U^{\delta,\eps}_i$, $i=1,2$, contains initial  conditions of motions for  which  change of  the action $I$  on the time interval $0\le \tau\le K$ tends to 0 as  $\eps\to 0$ (i.e ${\rm sup}_{0\le t\le K/\eps}\,|I(t)-I(0)|\to 0$ as $\eps\to 0$) and the phase point is in the domain $D_i$ at $\tau=K$.   It follows from results of \cite {N2} that ${\rm mes} \ v_{\eps} \to 0$ as  $\eps\to 0$. I.e. the adiabatic approximation works for the majority of initial conditions. Following \cite {A63} we call
 \begin{equation}
 \label{pr_hat}
 {\rm Pr}_{1,i}(  M)=\lim_{\delta\to 0}\lim_{\eps\to 0}\frac{{\rm mes}U^{\delta,\eps}_i}{{\rm mes}U^{\delta}}
 \end {equation} 
  the probability\footnote {This is the probability density for capture into $D_i$.} of capture of $  M$ into domain  $D_i$, $i=2,3$. This probability is a function of $\tau_*$, which we denote $P_{1,i}(\tau_*$),  ${\rm Pr}_{1,i}(  M)=P_{1,i}(\tau_*)$ \cite {N2}.  The phase portrait of the unperturbed system Fig. \ref{crossing} is different from that in \cite {N2}. Nevertheless it is possible to use an approach in \cite {N2} to formulate a procedure for calculation of $P_{1,i}(\tau_*)$  as follows.  
  
 Phase points from $U^{\delta}$ make many rounds repeatedly crossing the line $l=\pi$ in $D_1$ and gradually approaching separatrices. At each such round  near the separatrices the value of $E$ decays by about  $2\varepsilon\Theta$. So, at the last arrival  to the line $l=\pi$ in $D_1$  we have $0< E< 2\varepsilon\Theta$ (in the principal approximation). Phase points from $U^{\delta}$ finish  this last round almost simultaneously at $\tau \approx  \tau_*$. Consider motion of a phase point that starts at $\tau=\tau_*$ from the line $l=\pi$ with $E=E_*$, where $0<E_*< 2\varepsilon\Theta_*$,  $\Theta_*=\Theta(\tau_*$). These phase point will  make several rounds near separatrices.  The change of energy for motion near separatrices will be calculated in the principal approximation by formulas (\ref{p_delta}) with $\Theta=\Theta_*,\  \rho=\rho_*$.
 This construction  determines parts of the interval $0<E_*< 2\varepsilon\Theta_*$ corresponding to captures into $D_2$ and $D_3$.
 The ratio  to  $2\varepsilon\Theta_*$  of  length of the part, corresponding to capture into domain $D_i$, is equal to the probability    $P_{1,i}(\tau_*)$.
 
 Calculate probabilities using such an approach. We will assume that  $\rho_*>0$. The situation with $\rho_*<0$ is reduced to that with $\rho_*>0$ by exchange of numbers of domains  $D_2$ and $D_3$. 
 
 \medskip
  We will consider two cases.

{\bf Case I:  $\Theta_*>2\rho_*$}. A phase point  starts at $\tau=\tau_*$ from the line $l=\pi$ with $E=E_*$,  $0< E_*< 2\varepsilon\Theta_*$.  It passes near the separatrix $\Gamma_{1,3}$ during some time interval $\tau_*\le \tau\le \tau_1$ first, and arrives either to the line $L=0$ near $l=0$ in $D_3$ or to the line $l=0$ in $D_1$.   At the end of this pass the value of $E$ is $E_1= E_*-\eps(\Theta_* -2\rho_*)$.  Two sub-cases are possible.

 a) If $E_1<0$, then at $\tau=\tau_1$  the phase point is in $D_3$. In further motion it will make rounds near $\Gamma_{4,3}\cup\Gamma_{1,3}$, returning to the line $L=0$  near $l=0$ in $D_3$ after each round.   The value of $E$ decays by $2\eps\Theta_* $ at each such round.  This is a capture into  $D_3$.
 
  b) If $E_1>0$, then at $\tau=\tau_1$  the phase point is in $D_1$. Then it passes near the separatrix $\Gamma_{1,2}$ during some time interval $\tau_1\le \tau\le \tau_2$, and arrives  to the line $L=0$ near $l=\pi$  in $D_2$  with $E=E_2=E_1- \eps(\Theta_*+2\rho_*)=E_*-
  2\eps\Theta_*<0$.  In further motion it will make rounds near $\Gamma_{4,2}\cup\Gamma_{1,2}$ returning to $L=0$ near $l=\pi$ in $D_2$  after each round. The value of $E$ decays by $2\eps\Theta_* $ at each such round.  This is a capture into   $D_2$. 
  
  Thus, if  $\Theta_*>2\rho_*$, then  the probabilities 
are \begin{equation}
P_{1,2}=\frac{\Theta_*+2\rho_*}{2\Theta_*},\ \ P_{1,3}=\frac{\Theta_*-2\rho_*}{2\Theta_*}.
\end{equation}

{\bf Case II:  $\Theta_*<2\rho_*$}.  Let  $k$  be the natural number such that   $(2k-1)\Theta_*<2\rho_*<  (2k+1)\Theta_* $. A phase point  starts at $\tau=\tau_*$ from the line $l=\pi$ with $E=E_*$,  $0< E_*< 2\varepsilon\Theta_*$.  During a time interval $\tau_*\le \tau\le \tau_1$,  it goes along a path which is close to sequence of  segments   $\Gamma_{1,3}\cup\Gamma_{1,2}$ and $\Gamma_{4,2}\cup\Gamma_{4,3}$. The path contains $k$ such segments. At  $\tau=\tau_1$ the phase point is at the line $L=0$ near $l=\pi$  in $D_2$ or in $D_3$ with $E=E_1=E_*-2k\eps \Theta_*$. It is in $D_2$ if $k$ is odd, and in $D_3$, if $k$ is even. Consider, for definiteness, the case when $k$ is odd. The phase point passes near the separatrix $\Gamma_{4,2}$ during some time interval $\tau_1\le \tau\le\tau_2$. At the end of this pass the value of $E$ is $E_2= E_1-\eps(\Theta_* -2\rho_*)= E_*-(2k+1)\eps \Theta_* + 2\eps\rho_*$.  Two sub-cases are possible.

a)  If $E_2<0$, then at $\tau=\tau_2$  the phase point is in $D_2$. In further motion it will make rounds near $\Gamma_{1,2}\cup\Gamma_{4,2}$, returning to $L=0$ near $l=0$ in $D_2$  after each round.   The value of $E$ decays by $2\eps\Theta_* $ at each such round.  This is a capture into  $D_2$.

b) If $E_2>0$, then at $\tau=\tau_2$  the phase point is in $D_4$. Then it passes near the separatrix $\Gamma_{4,3}$ during some time interval $\tau_2\le \tau\le \tau_3$, and arrives  to the line $L=0$ near $l=\pi$ in $D_3$  with $E=E_3=E_2- \eps(\Theta_*+2\rho_*)=E_*-
  (2k+2) \eps\Theta_*<0$.  Then it passes near the separatrix $\Gamma_{1,3}$ during some time interval $\tau_3\le \tau\le \tau_4$.  At the end of this path $E=E_4 = E_3-\eps(\Theta_*-2\rho_*)=E_*-
  (2k+3) \eps\Theta_*+2\eps \rho_* < 2\eps\Theta_* - (2k+3) \eps\Theta_*+2\eps \rho_*= - (2k+1) \eps\Theta_*+2\eps \rho_*<0$.Thus at $\tau=\tau_4$ the phase point is at the line $L=0$ near $l=0$ in $D_3$.   In further motion it will make rounds near $\Gamma_{4,3}\cup\Gamma_{1,3}$ returning to $L=0$ near $l=0$  in $D_3$  after each round. The value of $E$ decays by $2\eps\Theta_* $ at each such round.  This is a capture into   $D_3$. 
  
Thus if  $(2k-1)\Theta_*<2\rho_*<  (2k+1)\Theta_* $ with an odd $k$, then  the probabilities 
are
\begin{equation}
P_{1,2}=\frac{(2k+1)\Theta_*-2\rho_*}{2\Theta_*},\ \ P_{1,3}=\frac{-(2k-1)\Theta_*+2\rho_*}{2\Theta_*}.
\end{equation}

Similar reasoning shows  that if $(2k-1)\Theta_*<2\rho_*<  (2k+1)\Theta_* $ with an even $k$, then  the probabilities are

\begin{equation}
P_{1,2}=\frac{-(2k-1)\Theta_*+2\rho_*}{2\Theta_*},\ \ P_{1,3}=\frac{(2k+1)\Theta_*-2\rho_*}{2\Theta_*}.
\end{equation}

Probabilities for initial conditions from $D_3$ can be obtained from the previous formulas by replacement of the index 1 with the index 4 and exchange of indexes 2 and 3.

\section*{Conclusion}
 We have described  evolution of  rotational dynamics of  a body with a slowly varying geometry of masses using an adiabatic approximation. The separatrix crossing in the course of this evolution is associated with a probabilistic scattering of phase trajectories.   We have calculated  probabilities of different outcomes of the evolutions   due to this scattering. These results could be useful in study of rotation of celestial bodies.
 
 \section*{Acknowledgment}
 The authors are thankful to A.V.Bolsinov for useful discussions.
% \section*{Appendix 1. Calculation of function $H_1$}

%\begin{pro}\label{probability}
%\ \\
%1. Under the condition that $-\Theta+2\rho<0$, we have the probabilities:
%$$P_{1,2}=\frac{\Theta+2\rho}{2\Theta},\ \ P_{1,3}=\frac{\Theta-2\rho}{2\Theta}.$$
%2. Under the condition that $-\Theta+2\rho>0$, and  $-(2k-1)\Theta+2\rho>0$, $-(2k+1)\Theta+2\rho<0$, $k\in\mathbb{N}$, we have the probabilities:
%\begin{equation}
%\left\{
%\begin{split}
%&P_{1,2}=\frac{(2k+1)\Theta-2\rho}{2\Theta},\ \ P_{1,3}=1-P_{1,2}=\frac{-(2k-1)\Theta+2\rho}{2\Theta},\ \ k\ is\ odd\\
%&P_{1,2}=\frac{-(2k-1)\Theta+2\rho}{2\Theta},\ \ P_{1,3}=1-P_{1,2}=\frac{(2k+1)\Theta-2\rho}{2\Theta},\ \ k\ is\ even.\\
%\end{split}
%\right.
%\nonumber
%\end{equation}
%\end{pro}

\end{document}